\documentclass[aps,twocolumn,superscriptaddress,floatfix,noeprint, pra]{revtex4-2}
\usepackage{float}
\usepackage{algorithm}
\usepackage{algpseudocode}
\usepackage{blindtext}
\usepackage{bm}
\usepackage{braket}
\usepackage{amssymb}
\usepackage{amsmath}
\usepackage{graphicx}
\usepackage{float}
\usepackage{bbold}
\usepackage{xcolor}
\usepackage{mathtools}
\usepackage{mathrsfs}
\usepackage[normalem]{ulem}
\usepackage[breaklinks=true,colorlinks=true,linkcolor=teal,urlcolor=teal,citecolor=teal]{hyperref}

\algrenewcommand\algorithmicrequire{\textbf{Input:}}
\algrenewcommand\algorithmicensure{\textbf{Output:}}
\newcommand{\I}{\mathrm{i}}
\newcommand{\id}{\mathbb{1}}

\renewcommand{\t}[1]{\textrm{#1}}

\newcommand{\mat}[2]{\left[ \begin{array}{#1} #2  \end{array}\right]}
\newcommand{\cov}{\Sigma}
\begin{document}
\title{Asymptotically optimal joint phase and dephasing strength estimation using spin-squeezed states}
\author{Arkadiusz Kobus}
\affiliation{Faculty of Physics, University of Warsaw, Pasteura 5, 02-093 Warszawa, Poland}
\affiliation{Institute of Physics PAS, Aleja Lotników 32/46, 02-668 Warszawa, Poland}
\author{Rafa{\l}  Demkowicz-Dobrza{\'n}ski} 
\affiliation{Faculty of Physics, University of Warsaw, Pasteura 5, 02-093 Warszawa, Poland}

\begin{abstract}
We show an explicit $N$-qubit protocol involving one-axis-twisted spin squeezed states, that allows for simultaneous phase and dephasing strength estimation with precision that asymptotically matches fundamental quantum metrological bounds. The relevance of the protocol goes beyond this particular model, since any uncorrelated noise quantum metrological model, that allows for at most constant asymptotic quantum enhancement, can be reduced to this problem via an appropriately tailored quantum error-correction procedure. 
\end{abstract}

\maketitle

\section{Introduction}
Quantum phase estimation (QPE) is a paradigmatic quantum estimation problem that captures the essence of the potential quantum enhancements in quantum metrology  \cite{Giovaennetti2006, Toth2014, Demkowicz2015, PezzeSmerziOberthalerEtAl2018, Polino2020}.
Use of entangled states, or coherent sequential  protocols allows for a quadratic scaling improvement in QPE precision, 
referred to as the \emph{Heisenberg scaling} (HS) 
\cite{Holland1993, Ou1997, Higgins2007, Pezze2008, Gorecki2020}. 

In realistic scenarios, however, the presence of noise results in decoherence of quantum probes typically prevents the HS to persist, in the asymptotic limit of large number of sensing probes or sequential probing channel uses, and reduces the effective quantum enhancement to a constant factor gain \cite{Escher2011, Demkowicz2012, Kolodynski2013, Demkowicz2014, Demkowicz2017, Zhou2018, Zhou2021}. 

The most common decoherence process that significantly affects the performance of phase estimation protocols, and in particular invalidates the asymptotic HS, is  \emph{dephasing}. The dephasing process causes loss of quantum coherence in superposition states that are used in any phase estimation protocol. In some sense, the dephasing process may be regarded as the canonical adversary for the QPE, as it targets the exact same quantum features that are crucial for any quantum enhanced phase estimation protocol to work. On a positive side, the optimal constant factor quantum enhancement that is achievable in the \emph{quantum phase estimation in presence of dephasing} (QPED) model \cite{Huelga1997, Escher2011, Demkowicz2012}, may be asymptotically saturated using a relatively simple strategy that utilizes \emph{one-axis twisted} (OAT) spin-squeezed states and a standard Ramsey interferometry protocol \cite{Orgikh2001, Ma2011, Escher2011}.  

Crucially, in the light of latest theoretical developments in quantum metrology,  QPED is not just
\emph{one of many} quantum metrological models
but in fact stands at the heart of the theory. 
In \cite{Zhou2020a, Zhou2021} it was shown, that all 
single-parameter quantum metrological models with parameter $\theta$ to be estimated, where the noise affects the sensing probes in a uncorrelated (Markovian) way, may be mapped via an appropriate quantum error-correction (QEC) protocols to 
either a noise-less QPE problem (in which case HS may be asymptotically preserved) or
to the QPED model (in which case only constant factor improvement is asymptotically possible).
Moreover, the mapping is asymptotically optimal, in the sense that the resulting protocols achieve the values of quantum Fisher information (QFI)  that coincide in the leading order with the ones predicted by fundamental bounds \cite{Zhou2021}.

Interestingly, the resulting QPED model, that is obtained via the above QEC mapping, will in general have both the phase $\varphi$ and the dephasing strength $\eta$ dependent on the parameter $\theta$.  
Therefore, in order to extract the complete information on the original parameter $\theta$ one needs to perform a \emph{simultaneous} phase and dephasing 
strength estimation in the optimal way.   
In the original paper \cite{Zhou2021} a general argument is provided that using spin-squeezed states it is possible to extract this information. However, the argument does not provide an explicit measurement protocols that achieves this. 

The above mentioned results are crucial for the whole theory as they at the same time prove the universal asymptotic saturabilty of quantum metrological bounds in presence of uncorrelated noise \cite{Escher2011, Demkowicz2012, Kolodynski2013, Demkowicz2014, Demkowicz2017, Zhou2018} as well as asymptotic equivalence of parallel and adaptive strategies \cite{Kurdzialek2023}. 
Hence, the protocol we describe in this paper is the last missing component in this impressive theoretical construction, that allows for an explicit universal construction of asymptotically optimal quantum metrological protocols.

The protocol utilizes the popular OAT states, this time with a twist, that the states must allow for optimal simultaneous estimation of \emph{both} the phase and the dephasing strength. This imposes non-trivial constraint, on the scaling of the squeezing strength parameter with the number of particle used, which is more stringent that in the case where only the phase is to be estimated. 
Moreover, the protocol requires the measurement of total angular momentum $J^2$, in addition to a single total angular momentum component $J_y$ which is measured in the standard Ramsey interferometry scenario. 

We should stress that the problem we address in this paper is qualitatively different that the one analyzed in e.g. \cite{Vidrighin2014, Len2022}, where also the tasks of simultaneous phase and dephasing strength estimation have been considered. In the above-mentioned papers, the focus was on the measurement incompatibility aspect of the problem, and how the incompatibility fades away with the increasing number of qubits involved. Unlike in our paper, though, the state of the qubits considered was simply a multiple-copy product state. As such, it was not the state that offers the optimal phase estimation performance. Hence, these results are not helpful, in completing the explicit construction of the universal QEC + OAT strategy that provides asymptotically optimal performance for any uncorrelated-noise quantum metrological models and the protocol we propose here is required.

\section{Metrological model}
Let us introduce a qubit quantum channel, that involves a rotation of the qubit state by a phase $\varphi \in [0,2\pi[$ around the $z$ axis in the Bloch sphere representation, accompanied by a dephasing process 
that shrinks the Bloch vector by a factor $\eta \in [0,1]$ in the $xy$ plane:
\begin{equation}
\label{eq:channel}
\Lambda_{\eta\varphi}(\rho) = 
U_\varphi \left(\sum_{k=0}^1 K_{\eta}^{(k)}\rho 
K_{\eta}^{(k)\dagger}\right) U_\varphi^\dagger, \quad U_\varphi = e^{-\frac{\I}{2} \varphi \sigma_z},
\end{equation}    
where Kraus operators representing dephasing are given by:
\begin{equation}
    K_\eta^{(0)}= \sqrt{\frac{1+\eta}{2}} \id, \quad K_{\eta}^{(1)} = \sqrt{\frac{1-\eta}{2}} \sigma_z.
\end{equation}
Given a general input state $\rho$, the output state 
$\rho_{\eta\varphi} = \Lambda_{\eta\varphi}(\rho)$ in the standard qubit basis reads:
\begin{equation}
\rho = \mat{cc}{\rho_{00} & \rho_{01} \\ \rho_{10} & \rho_{11}}  \overset{\Lambda_{\eta\varphi}}{\mapsto} \rho_{\eta\varphi} = \mat{cc}{\rho_{00} & \rho_{01}\eta e^{-\I \varphi} \\ \rho_{10} \eta e^{\I \varphi}& \rho_{11}},
\end{equation}
which manifest reduction of off-diagonal terms (loss of coherence) by a factor $\eta$---note that due to $z$-axis rotational symmetry of the dephasing process, the order in which the unitary rotation and the dephasing are applied is irrelevant.

In order to estimate parameters $\eta$, $\varphi$ one should measure the output state $\rho_{\eta\varphi}$, and 
try to infer the values of the parameters based on the measurement outcomes. 
Clearly, a single-qubit measurement will not be sufficient to get a meaningful estimation, and one should repeat the measurement procedure multiple times in order to gather enough information so that the parameters may be estimated with high precision. A quantitative bound on the achievable estimation precision  follows from the (multi-parameter) quantum Cram{\'e}r-Rao (QCR) matrix inequality \cite{Liu2019}:  
\begin{equation}
\label{eq:qfi}
    \cov\left(\tilde{\eta},\tilde{\varphi} \right)  \succeq \frac{1}{N} F_Q^{-1},  \quad (F_Q)_{\alpha \beta} = \t{Re} \left[\t{Tr}\left(\rho_{\eta\varphi} L_\alpha L_\beta\right)\right],
\end{equation}
where 
\begin{equation}
\cov(\tilde{\eta},\tilde{\varphi}) = \mat{cc}{\t{Var} (\tilde{\eta}) &  \t{Cov}(\tilde{\eta},\tilde{\varphi})\\\t{Cov}(\tilde{\eta},\tilde{\varphi})  & \t{Var} (\tilde{\varphi}) }
\end{equation}
is the covariance matrix of (locally unbiased) estimators $\tilde{\eta}$, $\tilde{\varphi}$, $N$ is the number of experiment repetitions, 
$F_Q$ is the QFI matrix with indices $\alpha,\beta \in \{\eta,\varphi\}$, whereas $L_\alpha$ are symmetric logarithmic derivative (SLD) operators, defined implicitly via:
\begin{equation}
    \partial_\alpha \rho_{\eta\varphi} = \frac{1}{2} \left(\rho_{\eta\varphi} L_\alpha + L_\alpha  \rho_{\eta\varphi} \right).
\end{equation}
The SLD operator $L_\alpha$ also contains information about the optimal measurement, as it is diagonal in the measurement basis saturating single-parameter QCR bound $\t{Var}(\tilde \alpha)=(F_Q)_{\alpha\alpha}^{-1}$.

One may now use the QFI matrix, to quantify the precision with which one can estimate a given parameter encoded in a quantum state. In the model considered, it is not difficult to show \cite{Vidrighin2014}, that the optimal input state $\rho$ that yields the output state $\rho_{\eta\varphi}$ for which the QFI matrix is the largest (in the matrix sense) is any state on the Bloch sphere equator, e.g. $\rho=\ket{+}\!\bra{+}$, $\ket{+}= \frac{1}{\sqrt{2}}(\ket{0} + \ket{1})$.
From \eqref{eq:qfi}, the corresponding QFI matrix reads:
\begin{equation}
    F_Q = \mat{cc}{\frac{1}{1-\eta^2} & 0 \\ 
    0 & \eta^2},  \ L_\eta=\frac{1}{1-\eta^2}(\sigma_x - \eta \id), \ L_\varphi= \eta \sigma_y,
\end{equation}
where we also provided explicit forms of the SLD operators, assuming for simplicity that the estimation is performed around point $\varphi=0$.
In this case, there is no probe-incompatibility issue, as there is a single input state that provides maximal QFI for each of the parameters simultaneously \cite{Ragy2016, Albarelli2022}. However, measurement incompatibility is present, 
and there is no single measurement that would make the QCR bound \eqref{eq:qfi} saturable \cite{Vidrighin2014}. This is related with the fact that the respective SLDs do not commute, and therefore a basis diagonalising both SLDs simultaneously does not exist. In order to extract maximal information on $\eta$ one should perform a measurement in the $\sigma_x$ eigen basis (optimal sensitivity for shrinking of the Bloch vector), while in order to extract the optimal information on $\varphi$, one should measure $\sigma_y$. As qubit bases can be viewed as directions on Bloch sphere, we could guess the optimal measurements based on geometry---the $x$ axis is the $\eta$-coordinate curve for $\varphi=0$, while the tangent to $\varphi$-coordinate curve for any $\eta$ and $\varphi=0$ is parallel to the $y$ axis.

For the multi-copy case with $N$ repetitions, we treat system of $N$ probe qubits as spin-$1/2$ particles, and introduce the collective angular momentum operators:
\begin{equation}
\label{eq:jcollective}
    J_i = \frac{1}{2}\sum_{n=1}^N \sigma_i^{(n)}, \quad i \in \{x,y,z\},
\end{equation}
where $\sigma_i^{(n)}$ represents a Pauli $\sigma_i$ operator acting on the $n$-th particle and the identity operator on the remaining subsystems. To remain within regime of independent repetitions, we consider only the product input states and local measurements, and via the previous single-copy considerations we obtain the optimal input state $\rho^{(N)} = \ket{+}\!\bra{+}^{\otimes N}$ along with measurements $J_x$ and $J_y$---optimal for $\eta$ and $\varphi$ estimation respectively.

Interestingly the measurement incompatibility may be avoided in the multi-copy limit, if one allows for collective measurements on all the $N$-qubit states 
$\rho_{\eta\varphi}^{\otimes N}$ \cite{Vidrighin2014, DemkowiczDobrzanski2020}. This is related with the fact that even though $[L_\eta,L_\varphi]\propto J_z\neq 0$, 
we have $\t{Tr}(\rho_{\eta\varphi}[L_\eta,L_\varphi])=0$, which is an if and only if condition for asymptotic saturability of the multiparameter QCR bound \cite{Crowley2014,  Ragy2016, Suzuki2016, DemkowiczDobrzanski2020}. This result may also be understood from the Quantum Local Asymptotic Normality (QLAN) perspective \cite{Kahn2009, DemkowiczDobrzanski2020}, as
the phase and the length of the Bloch vector are asymptotically mapped to quantum and classical degrees of freedom respectively, which results in no measurement-incompatibility.

Hence, for $N$ qubit product input state we may achieve asymptotically:
\begin{equation}
\cov(\tilde{\eta},\tilde{\varphi}) \underset{\t{product}}{\overset{N \rightarrow \infty}{\longrightarrow}} \frac{1}{N}\mat{cc}{1-\eta^2 & 0 \\ 0 & \frac{1}{\eta^2}}. 
\end{equation}

In general, however, product state strategies are not optimal. In presence of dephasing one can prepare entangled state of $N$ qubits that asymptotically achieves \cite{Orgikh2001, Ma2011, Escher2011}
\begin{equation}
    (F_Q)^{(N)}_{\varphi\varphi} = \frac{N\eta^2}{1-\eta^2}>N\eta^2=N(F_Q)_{\varphi\varphi},
\end{equation}
which is a factor of $1/(1-\eta^2)$ larger than for the the product case. This value coincides with fundamental metrological bounds, and as a result this value cannot be improved even with the help of the most general adaptive strategies involving QEC \cite{Demkowicz2014}. 
Moreover, the asymptotic optimal performance may be achieved with a simple Ramsey interferometric strategy involving OAT states \cite{Orgikh2001, Ma2011}. The gap between optimal precision and one obtained using product input state can be understood at the level of linear error propagation formula:
\begin{equation}
    \t{Var}(\tilde \varphi)=\frac{\t{Var}(J_y)}{\left|\frac{\partial\langle J_y\rangle}{\partial\varphi}\right|^2},
\end{equation}
where expectation values are computed on the output state $\rho_{\eta\varphi}^{(N)}$. While product input state maximises the denominator $\left|\frac{\partial\langle J_y\rangle}{\partial\varphi}\right|^2$, it provides sub-optimal value of the numerator $\t{Var}(J_y)$. The OAT states can asymptotically minimise the value of $\t{Var}(J_y)$ on the input state, which also decreases the resulting variance on the output state, and subsequently reduces estimation error $\t{Var}(\tilde \varphi)$.

In case of $\eta$ estimation the situation is simpler, as the product input state $\ket{+}\!\bra{+}^{\otimes N}$ is in fact the optimal state and the resulting QFI coincides with the fundamental bound, $(F_Q)^{(N)}_{\eta\eta} = N/(1-\eta^2)$ \cite{Fujiwara2003, Kolodynski2013}. 

Consequently, the fundamental bound on the covariance matrix for the simultaneous estimation of $\varphi$ and $\eta$, in the most general quantum metrological protocol involving $N$ uses of the channel \eqref{eq:channel}, reads:
\begin{equation}
\label{eq:limit qfi}
    \cov(\tilde{\eta},\tilde{\varphi}) \underset{\t{general}}{\succeq} 
    \frac{1}{N}\mat{cc}{1-\eta^2 & 0 \\ 
    0 & \frac{1-\eta^2}{\eta^2}}.
\end{equation}
In the next section we show an explicit protocol utilizing OAT states, which asymptotically saturates the above fundamental bound. 
Unlike in the Ramsey interferometric protocol that is optimal when estimating $\varphi$ only, we will need to measure two observables $J^2$ and $J_y$ simultaneously and will have to satisfy more stringent requirements regarding the scaling of the squeezing strength with $N$.

\section{Asymptotically optimal protocol}
\begin{figure*}[t]
    \centering
    \includegraphics[width=\textwidth]{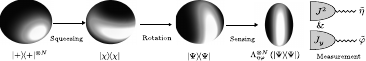}
    \caption{Schematic representation of the protocol. First the OAT state is prepared via a squeezing transformation. Then, it is rotated so that the minimal variance direction is along the $y$ axis. Sensing rotates the state by an angle $\varphi$ around the $z$ axis, as well as reduces the length of its angular momentum by approximately a factor of $\eta$. By performing a joint measurement of both $J^2$ and $J_y$ observables one may estimate $\eta$ and $\varphi$ simultaneously. If the squeezing parameter scales as $\chi \propto N^p$, $p \in ]-1,-3/4[$, the protocol asymptotically saturates the fundamental metrological bounds simultaneously for both $\eta$ and $\varphi$ estimation.}
    \label{fig:diagram}
\end{figure*}


The OAT $N$-qubit state \cite{Ma2011, PezzeSmerziOberthalerEtAl2018, Plodzien2022} is obtained through the application of squeezing (twisting) operation to a product state of equatorial qubit states:
\begin{equation}
\ket{\chi} = e^{-\I \chi J_z^2 }  \ket{+}^{\otimes N},
\end{equation}
where $\chi$ represents the strength of squeezing.
We follow the standard procedure \cite{Ma2011} and rotate the state around the $x$ axis, by an appropriately chosen angle so that the state has minimal variance of the $J_y$ angular momentum component, see Fig.~\ref{fig:diagram}. In this way we obtain the optimally \emph{rotated one-axis twisted} (ROAT) state:
\begin{equation}
\label{eq:epsilon}
    \ket{\Psi} = e^{\I J_x (\epsilon+\frac{\pi}{2})} \ket{\chi}, \ \epsilon = \frac{1}{2}\t{atan} \left(\frac{4 \sin \chi \cos^{N-2}\chi}{1-\cos^{N-2}2\chi} \right).
\end{equation}
The goal is to obtain a state with $J_y$ variance that scales more favourably with $N$ that in the case of the spin-coherent state $\ket{+}^{\otimes N}$, when it scales linearly $\t{Var}(J_y)_{\ket{+}^{\otimes N}} = N/4$. This is achieved whenever qubits are negatively correlated in the $y$ direction, which is indeed the case for ROAT, see Appendix \ref{sec:moment derivation}: 
\begin{equation}
\langle\sigma_y^{(m)}\sigma_y^{(n)}\rangle_\Psi<0,\  m\neq n.
\end{equation}
We assume power law scaling of squeezing strength $\chi \propto N^p$ with $p \in ]-1, -1/2[$, for which we can asymptotically expand ROAT moments,
leading to $J_y$ variance scaling sub-linearly with $N$:
\begin{equation}
\frac{\t{Var}(J_y)}{N} = O(N^2\chi^4,N^{-2}\chi^{-2}) \overset{N \rightarrow \infty}{\longrightarrow} 0.  
\end{equation}
The value of $p$ which asymptotically minimise $\t{Var}(J_y)$ can be found by fixing both error terms to be equal:
\begin{equation}
\label{eq:error terms}
    O(N^2\chi^4)\propto O(N^{-2}\chi^{-2})\iff {2+4p}={-2-2p},
\end{equation}
with $p=-2/3$ being the solution.

Such a prepared probe state is then subject to parallel action of $N$ parameters encoding channels $\Lambda_{\eta\varphi}$, resulting in the final mixed state:
\begin{equation}
    \rho^{(N)}_{\eta\varphi} = \Lambda_{\eta\varphi}^{\otimes N} \left( \ket{\Psi}\!\bra{\Psi}\right).
\end{equation}
Similarly as with product input state $\ket{+}\!\bra{+}^{\otimes N}$ we obtain the optimal estimation of $\eta$ or $\varphi$ with respectively $J_x$ or $J_y$ observable measurement on the final state. However, as $[J_x,J_y]\neq 0$, we cannot perform these measurements simultaneously. Focusing on the $J_x$ measurement, for product input state $\ket{+}\!\bra{+}^{\otimes N}$ it serves as the most informative measurement about the length of Bloch vector of the output state $\rho_{\eta\varphi}^{(N)}$---similarly with ROAT as the input state. Fortunately, we identify another observable---the total angular momentum $J^2$--- which can serve similar purpose in this scenario. Generally, $J^2$ measurement provides the information about spin alignment, and in our model all of the spin miss-alignment is generated by dephasing. Moreover, the $J^2$ operator \emph{do commute} with the $J_y$ operator, which means both can be measured simultaneously.

Since, we will be focusing on the $N \rightarrow \infty$ limit, we may assume that the estimated values will be converging to the true values, and in order to obtain the form of the estimators covariance matrix it is justified to use the multivariate linear error propagation formula:
\begin{align}
\label{eq:error propagation}
 \cov(\tilde{\eta},\tilde{\varphi}) &= \left[D^T \cov(J^2,J_y)^{-1} D \right]^{-1},\\
 D &= \mat{cc}{\frac{\partial \langle J^2 \rangle}{\partial \eta} & \frac{\partial \langle J^2 \rangle}{\partial \varphi}\\ \frac{\partial \langle J_y \rangle}{\partial \eta}& \frac{\partial \langle J_y \rangle}{\partial \varphi}},
\end{align}
where $\cov(J^2,J_y)$ is the covariance matrix of the observables $J^2$ and $J_y$ computed on the output state $\rho_{\eta\varphi}^{(N)}$, while $D$ is the relevant Jacobi matrix.

In order to compute the estimators covariance matrix we need to compute up to fourth-order moments of angular momentum operators (required for the $\t{Var}(J^2)$). Since collective angular momentum operators \eqref{eq:jcollective} are sums of local Pauli matrices, this implies we need to compute expectation values of up to four-fold tensor products of Pauli operators on the state. Fortunately, the initial state is permutationally invariant and so is the action of the parallel channels. Hence, the moments will not depend on the permutation of the qubit indices. In what follows we will use a notation:
\begin{equation}
\label{eq:tensor notation}
    \langle \sigma_i \rangle, \langle \sigma_i \otimes \sigma_j  \rangle,\langle \sigma_i \otimes \sigma_j \otimes \sigma_k \rangle, \langle \sigma_i \otimes \sigma_j \otimes \sigma_k \otimes \sigma_l \rangle 
\end{equation}
to represent expectation values of tensor products of respective Pauli matrices computed on  $\rho_{\eta\varphi}$, that act on \emph{different} particles without specifying the particle labels---it is assumed that operators acting on the remaining particles are identities. In particular, first and second moments of $J_i$ operator can be computed as 
\begin{equation}
\label{eq:expand}
    \langle J_i \rangle = \frac{N}{2} \langle \sigma_i \rangle,\  
    \langle J_i^2 \rangle  = \frac{N}{4} + 
    \frac{N(N-1)}{4} \langle \sigma_i \otimes \sigma_i \rangle, 
\end{equation}
where the first term results from the  $\sigma_i^2 = \id$ identity, whenever given $\sigma_i$ operator hits the same particle twice---see Appendix \ref{sec:moment derivation} for more general expression that are relevant for the results of this paper including up to fourth moments of angular momentum operators operators.

Moreover, it is convenient to use the Heisenberg picture to translate the expectation values computed on final state $\rho_{\eta\varphi}^{(N)}$ in terms of expectation values computed on ROAT input state $\ket{\Psi}$:
\begin{equation}
    \langle A \rangle = \langle A^{\t{H}} \rangle_{\Psi}, \quad A^{\t{H}} = \Lambda^*_{\eta\varphi}(A),
\end{equation}
where $A^{\t{H}}$ is the Heisenberg picture evolved observable $A$ and $\Lambda^*_{\eta\varphi}$ is the dual channel---a channel in which Kraus operators are replaced by their Hermitian conjugates.
To do so it is enough to evolve individual Pauli operators using the dual channel $\sigma_i^{\t{H}} = \Lambda^*_{\eta\varphi}(\sigma_i)$  which results in 
\begin{equation}
\label{eq:hp}
\mat{c}{\sigma_x^{\t{H}} \\ \sigma_y^{\t{H}} \\ \sigma_z^{\t{H}}} = 
\mat{ccc}{ \eta \cos\varphi& -\eta \sin \varphi&  0\\  \eta \sin \varphi &  \eta \cos\varphi& 0 \\0 & 0& 1} \mat{c}{\sigma_x \\ \sigma_y \\ \sigma_z}
\end{equation}
Consequently, using \eqref{eq:hp} together with \eqref{eq:expand}, we find that the expectation values of the observables that we measure in our protocol are: 
\begin{equation}
    \langle J_y \rangle  = \frac{N \eta}{2} \left[ \cos\varphi \langle \sigma_y \rangle_{\Psi} +  \sin \varphi \langle \sigma_x \rangle_{\Psi}\right]
\end{equation}
\begin{align}
    \langle J^2 \rangle  = \frac{3N}{4} + \frac{N(N-1)}{4} \big[&  \eta^2 \langle \sigma_x\otimes \sigma_x \rangle_{\Psi} + \\
    +&\eta^2 \langle \sigma_y\otimes \sigma_y \rangle_{\Psi} + 
    \langle \sigma_z\otimes \sigma_z \rangle_{\Psi} \big].\nonumber
\end{align}
The relevant expectation values of the ROAT state, see Appendix \ref{sec:moment derivation}, are:
\begin{align}
    \langle \sigma_x \rangle_{\Psi} & = 1  + O(N \chi^2), \quad 
    \langle \sigma_y  \rangle_{\Psi} = 0, \\
    \langle \sigma_x \otimes \sigma_x \rangle_{\Psi} & = 1 + O (N \chi^2), \quad
    \langle \sigma_y \otimes \sigma_y \rangle_{\Psi}  = O(N^{-1}), \end{align}
where we ignored $\langle \sigma_z \otimes \sigma_z \rangle_\Psi$ term as it does not contribute after the derivatives over the parameters are taken. This allows us to compute the Jacobi matrix (taking derivatives at $\varphi =0$) in the leading order:
\begin{equation}
D = \frac{N \eta}{2}\mat{cc}{ N [1 +O(N \chi^2)] &  0 \\ 
0 & 1 + O(N \chi^2) }.
\end{equation}
Importantly, the matrix is diagonal, which represents the fact that each observable is  sensitive to small changes of only one of the parameters of interest.

Finally, we compute the covariance matrix $\cov(J^2,J_y)$. 
After some straightforward but tedious calculations, the details of the computation are given in the Appendix \ref{sec:moment derivation}, the covariance matrix at $\varphi=0$ in the leading order is given by:
\begin{align}
\cov(J^2,J_y) &=\mat{cc}{\t{Var}(J^2) & 0 \\ 0 & \t{Var}(J_y)},\\
\t{Var}(J^2)&=\frac{N^3\eta^2(1-\eta^2)}{4}[1+O(N^3\chi^4)],\\
\t{Var}(J_y)&=\frac{N(1-\eta^2)}{4}[1+O(N^2\chi^4,N^{-2}\chi^{-2})].
\end{align}
As in the case of the Jacobi matrix, the covariance matrix $\cov(J^2,J_y)$ is diagonal, which implies that the resulting estimators covariance matrix $\cov(\tilde{\eta},\tilde{\varphi})$, calculated using the error propagation formula \eqref{eq:error propagation}, is also diagonal:
\begin{align}
\label{eq:optimal covariance}
\cov(\tilde{\eta},\tilde{\varphi})&=\mat{cc}{\t{Var}(\tilde{\eta}) & 0 \\ 0 & \t{Var}(\tilde{\varphi})},\\
\t{Var}(\tilde{\eta})&=\frac{1-\eta^2}{N}[1+O(N^3\chi^4)],\\
\t{Var}(\tilde{\varphi})&=\frac{1-\eta^2}{N\eta^2}[1+O(N\chi^2,N^{-2}\chi^{-2})].
\end{align}

Hence, we see that provided all higher-order terms vanish asymptotically, we  saturate the fundamental bound \eqref{eq:limit qfi}. The correction terms in the $\t{Var}(\tilde{\varphi})$ vanish for any squeezing strength scaling as $\chi\propto N^p$ for $p\in]-1,-1/2[$, with the fastest convergence for $p=-3/4$ (unlike in the noiseless scenario, when the fastest convergence is achieved for $p=-2/3$, as explained in \eqref{eq:error terms}). $p=-2/3$ also minimises $\t{Var}(J_y)$, nonetheless it is not optimal for estimation purposes). However, the correction terms in the $\t{Var}(\tilde{\eta})$ vanish asymptotically only for $p<-3/4$, excluding the optimal scaling for $\varphi$ estimation---we need to squeeze more gently, in order to keep the state closer to the product state structure, which is optimal for $\eta$ estimation.

Under this tighter constraint on squeezing strength, we achieve the asymptotically optimal joint estimation protocol:
\begin{equation}
    \cov(\tilde{\eta},\tilde{\varphi}) \underset{p\in]-1,-3/4[}{\overset{N \rightarrow \infty}{\longrightarrow}}\frac{1}{N}\mat{cc}{1-\eta^2 & 0 \\ 
    0 & \frac{1-\eta^2}{\eta^2}}.
\end{equation}

\begin{figure}
    \centering
    \includegraphics[width=\linewidth]{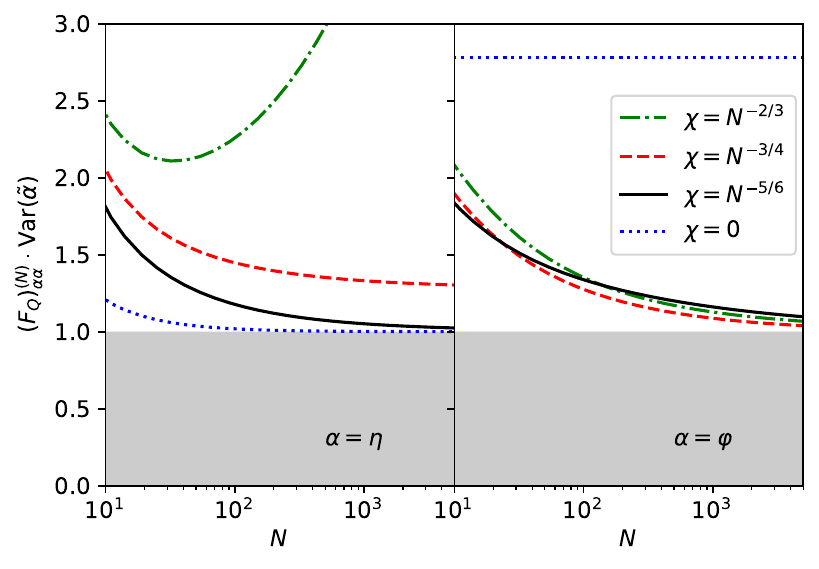}
    \caption{Variance of estimators $\tilde{\eta},\tilde{\varphi}$, normalised by the corresponding asymptotically optimal quantum Fisher information values, for different scalings of the squeezing strength $\chi$, and dephasing strength $\eta=0.8$. $\chi=N^{-2/3}$ scaling (dash-dotted green line) minimises $\t{Var}(J_y)$, $\chi=N^{-3/4}$ scaling (dashed red line) provides the fastest convergence to the asymptotic bound for $\varphi$ estimation, $\chi=0$ (blue dotted line) is optimal for $\eta$ estimation, while $\chi=N^{-5/6}$ (solid black line) allows for optimal simultaneous estimation with precisions converging to the fundamental bounds for both parameters. The gray area is excluded by the bounds.}
    \label{fig:precision plots}
\end{figure}

In order to obtain the fastest simultaneous convergence for both estimation variables, 
we are facing a trade-off between precision of estimating $\varphi$ (optimal at $p=-3/4$, with restriction on $p \in ]-1,-1/2[$ to converge to the fundamental bound) and $\eta$ (optimal at $p=-\infty$, with restriction $p \in [-\infty, -3/4[$ to converge to the fundamental bound). Analysis of the correction terms shows that the optimal choice is $p=-5/6$, see Fig.~\ref{fig:precision plots}.

\section{Summary}
We have presented a simple  protocol that provides optimal asymptotic performance for simultaneous phase and dephasing strength estimation. There are multiple experiments involving sensing with the help of OAT states \cite{PezzeSmerziOberthalerEtAl2018}, and so the protocol may be implemented using the state-of-the-art method, provided the squeezing strength is appropriately tuned to the number of atoms involved in the sensing procedure. 
The main challenge in the presented protocol is the measurement of $J^2$ observable in parallel with $J_y$. This requires exploiting some non-linear interaction between the subsystems which, however, are typically of the same nature as the interaction required to prepare squeezed states, and as such should be feasible as well with present-day experimental techniques.

This protocol, combined with QEC procedures that translate generic uncorrelated quantum metrological scenarios to the model considered \cite{Zhou2021}, completes the universal  construction of the asymptotic optimal quantum metrological protocol. Moreover, as we deal here with linear scaling of the relevant QFI, this procedure is also automatically asymptotically optimal from the Bayesian point of view \cite{Jarzyna2015}. 

\emph{Acknowledgments}. We thank Sisi Zhou for fruitful discussions. This work was supported by National Science Center (Poland) grant
No.2020/37/B/ST2/02134. 

\bibliography{phasedephase}

@Article{Jarzyna2015,
  author   = {Marcin Jarzyna and Rafał Demkowicz-Dobrzański},
  journal  = {New Journal of Physics},
  title    = {True precision limits in quantum metrology},
  year     = {2015},
  number   = {1},
  pages    = {013010},
  volume   = {17},
  url      = {http://stacks.iop.org/1367-2630/17/i=1/a=013010},
}

@Article{Plodzien2022,
  author    = {P\l{}odzie\ifmmode \acute{n}\else \'{n}\fi{}, Marcin and Lewenstein, Maciej and Witkowska, Emilia and Chwede\ifmmode \acute{n}\else \'{n}\fi{}czuk, Jan},
  journal   = {Phys. Rev. Lett.},
  title     = {One-Axis Twisting as a Method of Generating Many-Body Bell Correlations},
  year      = {2022},
  month     = {Dec},
  pages     = {250402},
  volume    = {129},
  doi       = {10.1103/PhysRevLett.129.250402},
  issue     = {25},
  numpages  = {7},
  publisher = {American Physical Society},
  url       = {https://link.aps.org/doi/10.1103/PhysRevLett.129.250402},
}

@article{Kahn2009,
  title = {Local Asymptotic Normality for Finite Dimensional Quantum Systems},
  volume = {289},
  ISSN = {1432-0916},
  url = {http://dx.doi.org/10.1007/s00220-009-0787-3},
  DOI = {10.1007/s00220-009-0787-3},
  number = {2},
  journal = {Communications in Mathematical Physics},
  publisher = {Springer Science and Business Media LLC},
  author = {Kahn,  Jonas and Gu\c{t}\u{a},  Mădălin},
  year = {2009},
  month = mar,
  pages = {597–652}
}

@Article{Suzuki2016,
  author  = {Suzuki,Jun},
  journal = {Journal of Mathematical Physics},
  title   = {Explicit formula for the Holevo bound for two-parameter qubit-state estimation problem},
  year    = {2016},
  number  = {4},
  pages   = {042201},
  volume  = {57},
  doi     = {10.1063/1.4945086},
  eprint  = {https://doi.org/10.1063/1.4945086},
  url     = {https://doi.org/10.1063/1.4945086},
}

@Article{Crowley2014,
  Title                    = {Tradeoff in simultaneous quantum-limited phase and loss estimation in interferometry},
  Author                   = {Crowley, Philip J. D. and Datta, Animesh and Barbieri, Marco and Walmsley, I. A.},
  Journal                  = {Phys. Rev. A},
  Year                     = {2014},
  Month                    = {Feb},
  Pages                    = {023845},
  Volume                   = {89},
  DOI                      = {10.1103/PhysRevA.89.023845},
  Issue                    = {2},
  Numpages                 = {9},
  Publisher                = {American Physical Society},
  URL                      = {https://link.aps.org/doi/10.1103/PhysRevA.89.023845}
}

@Article{Fujiwara2003,
  Title                    = {Quantum parameter estimation of a generalized Pauli channel},
  Author                   = {Akio Fujiwara and Hiroshi Imai},
  Journal                  = {J. Phys. A: Math. Gen.},
  Year                     = {2003},
  Pages                    = {8093-8103},
  Volume                   = {36}
}

@Article{DemkowiczDobrzanski2020,
  author    = {Rafał Demkowicz-Dobrzański and Wojciech Górecki and Mădălin Gu\c{t}\u{a}},
  journal   = {Journal of Physics A: Mathematical and Theoretical},
  title     = {Multi-parameter estimation beyond quantum Fisher information},
  year      = {2020},
  month     = {aug},
  number    = {36},
  pages     = {363001},
  volume    = {53},
  doi       = {10.1088/1751-8121/ab8ef3},
  publisher = {IOP Publishing},
  url       = {https://dx.doi.org/10.1088/1751-8121/ab8ef3},
}

@Article{Albarelli2022,
  author    = {Albarelli, Francesco and Demkowicz-Dobrza\ifmmode \acute{n}\else \'{n}\fi{}ski, Rafa\l{}},
  journal   = {Phys. Rev. X},
  title     = {Probe Incompatibility in Multiparameter Noisy Quantum Metrology},
  year      = {2022},
  month     = {Mar},
  pages     = {011039},
  volume    = {12},
  doi       = {10.1103/PhysRevX.12.011039},
  issue     = {1},
  numpages  = {28},
  publisher = {American Physical Society},
  url       = {https://link.aps.org/doi/10.1103/PhysRevX.12.011039},
}

@Article{Ragy2016,
  Title                    = {Compatibility in multiparameter quantum metrology},
  Author                   = {Ragy, Sammy and Jarzyna, Marcin and Demkowicz-Dobrza\ifmmode \acute{n}\else \'{n}\fi{}ski, Rafa\l{}},
  Journal                  = {Phys. Rev. A},
  Year                     = {2016},
  Month                    = {Nov},
  Pages                    = {052108},
  Volume                   = {94},
  DOI                      = {10.1103/PhysRevA.94.052108},
  Issue                    = {5},
  Numpages                 = {11},
  Publisher                = {American Physical Society},
  URL                      = {https://link.aps.org/doi/10.1103/PhysRevA.94.052108}
}

@Article{Liu2019,
  Title                    = {Quantum Fisher information matrix and multiparameter estimation},
  Author                   = {Jing Liu and Haidong Yuan and Xiao-Ming Lu and Xiaoguang Wang},
  Journal                  = {Journal of Physics A: Mathematical and Theoretical},
  Year                     = {2019},
  Month                    = {dec},
  Number                   = {2},
  Pages                    = {023001},
  Volume                   = {53},
  DOI                      = {10.1088/1751-8121/ab5d4d},
  Publisher                = {{IOP} Publishing},
  URL                      = {https://doi.org/10.1088%2F1751-8121%2Fab5d4d}
}

@article{Len2022,
doi = {10.1088/1367-2630/ac599d},
url = {https://dx.doi.org/10.1088/1367-2630/ac599d},
year = {2022},
month = {mar},
publisher = {IOP Publishing},
volume = {24},
number = {3},
pages = {033037},
author = {Len, Yink Loong},
title = {Multiparameter estimation for qubit states with collective measurements: a case study},
journal = {New Journal of Physics}
}

@Article{Vidrighin2014,
  Title                    = {{Joint estimation of phase and phase diffusion for quantum metrology}},
  Author                   = {{Vidrighin}, M.~D. and {Donati}, G. and {Genoni}, M.~G. and {Jin}, X.-M. and {Kolthammer}, W.~S. and {Kim}, M.~S. and {Datta}, A. and {Barbieri}, M. and {Walmsley}, I.~A.},
  Journal                  = {Nature Communications},
  Year                     = {2014},
  Month                    = apr,
  Pages                    = {3532},
  Volume                   = {5},
  DOI                      = {10.1038/ncomms4532},
  Eid                      = {3532}
}

@Article{Kolodynski2013,
  author   = {Jan Kolodynski and Rafal Demkowicz-Dobrzanski},
  journal  = {New Journal of Physics},
  title    = {Efficient tools for quantum metrology with uncorrelated noise},
  year     = {2013},
  number   = {7},
  pages    = {073043},
  volume   = {15},
  url      = {http://stacks.iop.org/1367-2630/15/i=7/a=073043},
}

@Article{Kurdzialek2023,
  author    = {Kurdzia\l{}ek, Stanis\l{}aw and G\'orecki, Wojciech and Albarelli, Francesco and Demkowicz-Dobrza\ifmmode \acute{n}\else \'{n}\fi{}ski, Rafa\l{}},
  journal   = {Phys. Rev. Lett.},
  title     = {Using Adaptiveness and Causal Superpositions Against Noise in Quantum Metrology},
  year      = {2023},
  month     = {Aug},
  pages     = {090801},
  volume    = {131},
  doi       = {10.1103/PhysRevLett.131.090801},
  issue     = {9},
  numpages  = {7},
  publisher = {American Physical Society},
  url       = {https://link.aps.org/doi/10.1103/PhysRevLett.131.090801},
}

@Article{Huelga1997,
  Title                    = {Improvement of Frequency Standards with Quantum Entanglement},
  Author                   = {Huelga, S. F. and Macchiavello, C. and Pellizzari, T. and Ekert, A. K. and Plenio, M. B. and Cirac, J. I.},
  Journal                  = {Phys. Rev. Lett.},
  Year                     = {1997},
  Month                    = {Nov},
  Number                   = {20},
  Pages                    = {3865--3868},
  Volume                   = {79},
  DOI                      = {10.1103/PhysRevLett.79.3865},
  Numpages                 = {3},
  Publisher                = {American Physical Society}
}

@Article{Orgikh2001,
  Title                    = {Spin squeezing and decoherence limit in Ramsey spectroscopy},
  Author                   = {Ulam-Orgikh, Duger and Kitagawa, Masahiro},
  Journal                  = {Phys. Rev. A},
  Year                     = {2001},
  Month                    = {Oct},
  Pages                    = {052106},
  Volume                   = {64},
  DOI                      = {10.1103/PhysRevA.64.052106},
  Issue                    = {5},
  Numpages                 = {6},
  Publisher                = {American Physical Society},
  URL                      = {http://link.aps.org/doi/10.1103/PhysRevA.64.052106}
}

@Article{Ma2011,
  author    = {Jian Ma and Xiaoguang Wang and C.P. Sun and Franco Nori},
  journal   = {Physics Reports},
  title     = {Quantum spin squeezing},
  year      = {2011},
  issn      = {0370-1573},
  number    = {2–3},
  pages     = {89 - 165},
  volume    = {509},
  doi       = {http://dx.doi.org/10.1016/j.physrep.2011.08.003},
  url       = {http://www.sciencedirect.com/science/article/pii/S0370157311002201},
}

@Article{Zhou2020a,
  Title                    = {Optimal approximate quantum error correction for quantum metrology},
  Author                   = {Zhou, Sisi and Jiang, Liang},
  Journal                  = {Phys. Rev. Research},
  Year                     = {2020},
  Month                    = {Mar},
  Pages                    = {013235},
  Volume                   = {2},
  DOI                      = {10.1103/PhysRevResearch.2.013235},
  Issue                    = {1},
  Numpages                 = {12},
  Publisher                = {American Physical Society},
  URL                      = {https://link.aps.org/doi/10.1103/PhysRevResearch.2.013235}
}

@Article{Zhou2021,
  author    = {Zhou, Sisi and Jiang, Liang},
  journal   = {PRX Quantum},
  title     = {Asymptotic Theory of Quantum Channel Estimation},
  year      = {2021},
  month     = {Mar},
  pages     = {010343},
  volume    = {2},
  doi       = {10.1103/PRXQuantum.2.010343},
  issue     = {1},
  numpages  = {25},
  publisher = {American Physical Society},
  url       = {https://link.aps.org/doi/10.1103/PRXQuantum.2.010343},
}

@Article{Zhou2018,
  author   = {Zhou, Sisi and Zhang, Mengzhen and Preskill, John and Jiang, Liang},
  journal  = {Nature Communications},
  title    = {Achieving the Heisenberg limit in quantum metrology using quantum error correction},
  year     = {2018},
  issn     = {2041-1723},
  month    = {Jan},
  number   = {1},
  pages    = {78},
  volume   = {9},
  day      = {08},
  doi      = {10.1038/s41467-017-02510-3},
  url      = {https://doi.org/10.1038/s41467-017-02510-3},
}

@Article{Demkowicz2017,
  Title                    = {Adaptive Quantum Metrology under General Markovian Noise},
  Author                   = {Demkowicz-Dobrza\ifmmode \acute{n}\else \'{n}\fi{}ski, Rafa\l{} and Czajkowski, Jan and Sekatski, Pavel},
  Journal                  = {Phys. Rev. X},
  Year                     = {2017},
  Month                    = {Oct},
  Pages                    = {041009},
  Volume                   = {7},
  DOI                      = {10.1103/PhysRevX.7.041009},
  Issue                    = {4},
  Numpages                 = {15},
  Publisher                = {American Physical Society},
  URL                      = {https://link.aps.org/doi/10.1103/PhysRevX.7.041009}
}

@Article{Demkowicz2014,
  Title                    = {Using Entanglement Against Noise in Quantum Metrology},
  Author                   = {Demkowicz-Dobrza\ifmmode \acute{n}\else \'{n}\fi{}ski, Rafal and Maccone, Lorenzo},
  Journal                  = {Phys. Rev. Lett.},
  Year                     = {2014},
  Month                    = {Dec},
  Pages                    = {250801},
  Volume                   = {113},
  DOI                      = {10.1103/PhysRevLett.113.250801},
  Issue                    = {25},
  Numpages                 = {5},
  Publisher                = {American Physical Society}
}

@Article{Escher2011,
  Title                    = {General framework for estimating the ultimate precision limit in noisy quantum-enhanced metrology},
  Author                   = {B. M. Escher and R. L. de Matos Filho and L. Davidovich},
  Journal                  = {Nature Phys.},
  Year                     = {2011},
  Pages                    = {406-411},
  Volume                   = {7},
  DOI                      = {10.1038/nphys1958}
}

@Article{Demkowicz2012,
  Title                    = {The elusive Heisenberg limit in quantum enhanced metrology},
  Author                   = {Demkowicz-Dobrza\'{n}ski, R. and Gu\c{t}\u{a}, M. and Ko{\l}ody\'{n}ski, J.},
  Journal                  = {Nat. Commun.},
  Year                     = {2012},
  Pages                    = {1063},
  Volume                   = {3}
}

@Article{PezzeSmerziOberthalerEtAl2018,
  Title                    = {Quantum metrology with nonclassical states of atomic ensembles},
  Author                   = {Pezz\`e, Luca and Smerzi, Augusto and Oberthaler, Markus K. and Schmied, Roman and Treutlein, Philipp},
  Journal                  = {Rev. Mod. Phys.},
  Year                     = {2018},
  Month                    = {Sep},
  Pages                    = {035005},
  Volume                   = {90},
  DOI                      = {10.1103/RevModPhys.90.035005},
  Issue                    = {3},
  Numpages                 = {70},
  Publisher                = {American Physical Society},
  URL                      = {https://link.aps.org/doi/10.1103/RevModPhys.90.035005}
}

@Article{Polino2020,
  Title                    = {{Photonic quantum metrology}},
  Author                   = {{Polino}, Emanuele and {Valeri}, Mauro and {Spagnolo}, Nicol{\`o} and {Sciarrino}, Fabio},
  Journal                  = {AVS Quantum Science},
  Year                     = {2020},
  Month                    = jun,
  Number                   = {2},
  Pages                    = {024703},
  Volume                   = {2},
  Adsurl                   = {https://ui.adsabs.harvard.edu/abs/2020AVSQS...2b4703P},
  Archiveprefix            = {arXiv},
  DOI                      = {10.1116/5.0007577},
  Eid                      = {024703},
  Eprint                   = {2003.05821},
  Primaryclass             = {quant-ph}
}

@Article{Ou1997,
  author    = {Ou, Z. Y.},
  journal   = {Phys. Rev. A},
  title     = {Fundamental quantum limit in precision phase measurement},
  year      = {1997},
  month     = {Apr},
  pages     = {2598--2609},
  volume    = {55},
  doi       = {10.1103/PhysRevA.55.2598},
  issue     = {4},
  numpages  = {0},
  publisher = {American Physical Society},
  url       = {https://link.aps.org/doi/10.1103/PhysRevA.55.2598},
}

@InCollection{Demkowicz2015,
  Title                    = {Quantum limits in optical interferometry},
  Author                   = {Demkowicz-Dobrzanski, R. and Jarzyna, M. and Ko\l{}ody\'{n}ski, J.},
  Booktitle                = {Progress in Optics},
  Publisher                = {Elsevier},
  Year                     = {2015},
  Editor                   = {Emil Wolf},
  Pages                    = {345--435},
  Volume                   = {60},
  DOI                      = {10.1016/bs.po.2015.02.003}
}

@Article{Toth2014,
  author    = {Geza Toth and Iagoba Apellaniz},
  journal   = {J. Phys. A: Math. Theor.},
  title     = {Quantum metrology from a quantum information science perspective},
  year      = {2014},
  number    = {42},
  pages     = {424006},
  volume    = {47},
  url       = {http://stacks.iop.org/1751-8121/47/i=42/a=424006},
}

@Article{Giovaennetti2006,
  Title                    = {Quantum Metrology},
  Author                   = {Giovannetti, Vittorio and Lloyd, Seth and Maccone, Lorenzo},
  Journal                  = {Phys. Rev. Lett.},
  Year                     = {2006},
  Month                    = {Jan},
  Pages                    = {010401},
  Volume                   = {96},
  DOI                      = {10.1103/PhysRevLett.96.010401},
  Issue                    = {1},
  Numpages                 = {4},
  Publisher                = {American Physical Society},
  URL                      = {http://link.aps.org/doi/10.1103/PhysRevLett.96.010401}
}

@Article{Gorecki2020,
  Title                    = {$\ensuremath{\pi}$-Corrected Heisenberg Limit},
  Author                   = {G\'orecki, Wojciech and Demkowicz-Dobrza\ifmmode \acute{n}\else \'{n}\fi{}ski, Rafa\l{} and Wiseman, Howard M. and Berry, Dominic W.},
  Journal                  = {Phys. Rev. Lett.},
  Year                     = {2020},
  Month                    = {Jan},
  Pages                    = {030501},
  Volume                   = {124},
  DOI                      = {10.1103/PhysRevLett.124.030501},
  Issue                    = {3},
  Numpages                 = {5},
  Publisher                = {American Physical Society},
  URL                      = {https://link.aps.org/doi/10.1103/PhysRevLett.124.030501}
}

@Article{Higgins2007,
  Title                    = {Entanglement-free Heisenberg-limited phase estimation},
  Author                   = {B.~L. Higgins and D.~W. Berry and S.~D. Bartlett and H.~M. Wiseman and G.~J. Pryde},
  Journal                  = {Nature},
  Year                     = {2007},
  Pages                    = {393},
  Volume                   = {450}
}

@Article{Holland1993,
  author    = {Holland, M. J. and Burnett, K.},
  journal   = {Phys. Rev. Lett.},
  title     = {Interferometric detection of optical phase shifts at the Heisenberg limit},
  year      = {1993},
  month     = {Aug},
  number    = {9},
  pages     = {1355--1358},
  volume    = {71},
  doi       = {10.1103/PhysRevLett.71.1355},
  numpages  = {3},
  publisher = {American Physical Society},
}

@Article{Pezze2008,
  author    = {Pezz\'e, Luca and Smerzi, Augusto},
  journal   = {Phys. Rev. Lett.},
  title     = {Mach-Zehnder Interferometry at the Heisenberg Limit with Coherent and Squeezed-Vacuum Light},
  year      = {2008},
  month     = {Feb},
  number    = {7},
  pages     = {073601},
  volume    = {100},
  doi       = {10.1103/PhysRevLett.100.073601},
  numpages  = {4},
  publisher = {American Physical Society},
}

\appendix

\section{Detailed derivation of moments}\label{sec:moment derivation}
We begin with calculation of OATSS qubit moments defined in \eqref{eq:tensor notation}. Note that the state $\ket{+}^{\otimes N}$ is invariant under rotation
\begin{equation}
\label{eq:symmetry}
   (J_x,J_y,J_z)\overset{e^{\I\pi J_x}}{\longrightarrow} (J_x,-J_y,-J_z),
\end{equation}
and so are states $U\ket{+}^{\otimes N}$ for any unitary $U$ that is also invariant under the this symmetry. In particular, both $\ket{\chi}=e^{-\I \chi J_z^2 } \ket{+}^{\otimes N}$ and $\ket{\Psi} = e^{\I J_x (\epsilon+\frac{\pi}{2})} \ket{\chi}$ are symmetric under this transformation, causing all moments with odd number of $\sigma_y$ and $\sigma_z$ operators to be identically zero for these states.

For computation of the rest of moments we evolve Pauli operators in the Heisenberg picture, following the procedure described in \cite{Ma2011}. We obtain the exact moments and expand them up to $O(N^2\chi^4)$ order for $p\in]-1,-1/2[$:
\begin{align*}
    \langle \sigma_x\rangle_\t{OAT}&=
    \cos^{N-1}\chi=1-\frac{1}{2}N\chi^2+{O}(N^2\chi^4),\\
    \langle \sigma_x\otimes\sigma_x\rangle_\t{OAT}&=
    \frac{1+\cos^{N-2}2\chi}{2}=1-N\chi^2+{O}(N^2\chi^4),\\
    \langle \sigma_y\otimes\sigma_y\rangle_\t{OAT}&=
    \frac{1-\cos^{N-2}2\chi}{2}= N\chi^2+{O}(N^2\chi^4),\\
    \langle \sigma_y\otimes\sigma_z\rangle_\t{OAT}&=
    \sin\chi\cos^{N-2}\chi=\chi+{O}(N\chi^3),\\
    \langle \sigma_z\otimes\sigma_z\rangle_\t{OAT}&=0,
\end{align*}
\begin{align*}
    \langle \sigma_x\otimes\sigma_x\otimes\sigma_x\rangle_\t{OAT}&=
    \frac{3\cos^{N-3}\chi+\cos^{N-3}3\chi}{4}=\\&=
    1-\frac{3}{2}N\chi^2+O(N^2\chi^4),\\
    \langle \sigma_x\otimes\sigma_y\otimes\sigma_y\rangle_\t{OAT}&=
    \frac{\cos^{N-3}\chi-\cos^{N-3}3\chi}{4}=\\&=
    N\chi^2+O(N^2\chi^4),\\
    \langle \sigma_x\otimes\sigma_y\otimes\sigma_z\rangle_\t{OAT}&=
    \frac{\sin2\chi\cos^{N-3}2\chi}{2}=\\&=
    \chi+O(N\chi^3),\\
    \langle \sigma_x\otimes\sigma_z\otimes\sigma_z\rangle_\t{OAT}&=
    \sin^2\chi\cos^{N-3}\chi=O(\chi^2),
\end{align*}
\begin{align*}
    \langle \sigma_x&\otimes\sigma_x\otimes\sigma_x\otimes\sigma_x\rangle_\t{OAT}=\\&=
    \frac{3+4\cos^{N-4}2\chi+\cos^{N-4}4\chi}{8}=\\&=
    1-2N\chi^2+{O}(N^2\chi^4),\\
    \langle \sigma_x&\otimes\sigma_x\otimes\sigma_y\otimes\sigma_y\rangle_\t{OAT}=
    \frac{1-\cos^{N-4}4\chi}{8}=\\&=
    N\chi^2+{O}(N^2\chi^4),\\
    \langle \sigma_x&\otimes\sigma_x\otimes\sigma_y\otimes\sigma_z\rangle_\t{OAT}=\\&=
    \frac{\sin\chi\cos^{N-4}\chi+\cos^{N-4}3\chi[3\sin\chi-4\sin^3\chi]}{4}=\\&=
    \chi+{O}(N\chi^3),\\
    \langle \sigma_x&\otimes\sigma_x\otimes\sigma_z\otimes\sigma_z\rangle_\t{OAT}=
    \frac{-\sin^22\chi\cos^{N-4}2\chi}{2}={O}(\chi^2),
\end{align*}
\begin{align*}
    \langle \sigma_y&\otimes\sigma_y\otimes\sigma_y\otimes\sigma_y\rangle_\t{OAT}=\\&=
    \frac{3-4\cos^{N-4}2\chi+\cos^{N-4}4\chi}{8}={O}(N^2\chi^4),\\
    \langle \sigma_y&\otimes\sigma_y\otimes\sigma_y\otimes\sigma_z\rangle_\t{OAT}=\\&=
    \frac{3\sin\chi\cos^{N-4}\chi-\cos^{N-4}3\chi[3\sin\chi-4\sin^3\chi]}{4}=\\&=
    {O}(N\chi^3),\\
    \langle \sigma_y&\otimes\sigma_y\otimes\sigma_z\otimes\sigma_z\rangle_\t{OAT}=
    \frac{\sin^22\chi\cos^{N-4}2\chi}{2}={O}(\chi^2),\\
    \langle \sigma_y&\otimes\sigma_z\otimes\sigma_z\otimes\sigma_z\rangle_\t{OAT}=
    -\sin^3\chi\cos^{N-4}\chi={O}(\chi^3),\\
    \langle \sigma_z&\otimes\sigma_z\otimes\sigma_z\otimes\sigma_z\rangle_\t{OAT}=0,
\end{align*}
and from this we also find ROAT moments \eqref{eq:epsilon}, after appropriately rotating OAT moments and expanding the angle of rotation $\epsilon$:
\begin{align*}
    \langle \sigma_x\rangle_\Psi&=1-\frac{1}{2}N\chi^2+{O}(N^2\chi^4),\\
    \langle \sigma_x\otimes\sigma_x\rangle_\Psi&=1-N\chi^2+{O}(N^2\chi^4),\\
    \langle \sigma_y\otimes\sigma_y\rangle_\Psi&=-N^{-1}+{O}(N\chi^4,N^{-3}\chi^{-2}),\\
    \langle \sigma_y\otimes\sigma_z\rangle_\Psi&=0,\\
    \langle \sigma_z\otimes\sigma_z\rangle_\Psi&=N\chi^2+N^{-1}+O(N^2\chi^4),\\
    \langle \sigma_x\otimes\sigma_x\otimes\sigma_x\rangle_\Psi&=1-\frac{3}{2}N\chi^2+O(N^2\chi^4),\\
    \langle \sigma_x\otimes\sigma_y\otimes\sigma_y\rangle_\Psi&=-N^{-1}+{O}(\chi^2,N^{-3}\chi^{-2}),\\
    \langle \sigma_x\otimes\sigma_y\otimes\sigma_z\rangle_\Psi&={O}(N\chi^3,N^{-2}\chi^{-1}),\\
    \langle \sigma_x\otimes\sigma_z\otimes\sigma_z\rangle_\Psi&=N\chi^2+N^{-1}+O(N^2\chi^4),
\end{align*}
\begin{align*}
    \langle \sigma_x\otimes\sigma_x\otimes\sigma_x\otimes\sigma_x\rangle_\Psi&=1-2N\chi^2+O(N^2\chi^4),\\
    \langle \sigma_x\otimes\sigma_x\otimes\sigma_y\otimes\sigma_y\rangle_\Psi&=-N^{-1}+{O}(\chi^2,N^{-3}\chi^{-2}),\\
    \langle \sigma_x\otimes\sigma_x\otimes\sigma_y\otimes\sigma_z\rangle_\Psi&={O}(N\chi^3,N^{-2}\chi^{-1}),\\
    \langle\sigma_x\otimes\sigma_x\otimes\sigma_z\otimes\sigma_z\rangle_\Psi&=N\chi^2+N^{-1}+O(N^2\chi^4),\\
    \langle \sigma_y\otimes\sigma_y\otimes\sigma_y\otimes\sigma_y\rangle_\Psi&={O}(N^{-2}),\\
    \langle \sigma_y\otimes\sigma_y\otimes\sigma_y\otimes\sigma_z\rangle_\Psi&={O}(\chi^3,N^{-3}\chi^{-1}),\\
    \langle \sigma_y\otimes\sigma_y\otimes\sigma_z\otimes\sigma_z\rangle_\Psi&={O}(\chi^2),\\
    \langle \sigma_y\otimes\sigma_z\otimes\sigma_z\otimes\sigma_z\rangle_\Psi&={O}(N^2\chi^5,N^{-1}\chi),\\
    \langle \sigma_z\otimes\sigma_z\otimes\sigma_z\otimes\sigma_z\rangle_\Psi&={O}(N^2\chi^4).
\end{align*}
Note, that the state $\ket{\Psi}$ is approximately Gaussian, i.e. satisfies following property:
\begin{align}
\label{eq:gaussian}
    \langle \sigma_i\otimes\sigma_i&\otimes\sigma_j\otimes\sigma_j\rangle_\Psi=\langle \sigma_i\otimes\sigma_i\rangle_\Psi\langle\sigma_j\otimes\sigma_j\rangle_\Psi+O(N^2\chi^4)
\end{align}
for all $i,j\in\{x,y,z\}$  with asymptotically vanishing error terms. We verify this by direct check of all possible cases. Notably error terms $O(N^{-3}\chi^{-2})$ in $\langle \sigma_y\otimes\sigma_y\rangle_\Psi$ perfectly cancels the error terms in $\langle \sigma_x\otimes\sigma_x\otimes\sigma_y\otimes\sigma_y\rangle_\Psi$, leaving only terms of order $O(\chi^2)$.

We proceed to derivation of the covariance matrix $\cov(J^2,J_y)$ for $\varphi=0$. The $\t{Cov}(J^2,J_y)=0$ due to the symmetry \eqref{eq:symmetry}. The $\t{Var}(J_y)$ follows directly from \eqref{eq:expand}:
\begin{align}
    \t{Var}(J_y)&=\frac{1}{4}[N+\eta^2N(N-1)\langle \sigma_y\otimes\sigma_y\rangle_\Psi]=\\&=\frac{N}{4}[1-\eta^2 + O(N^2\chi^4,N^{-2}\chi^{-2})].\nonumber
\end{align}
For the $\t{Var}(J^2)$ we use its decomposition:
\begin{equation}
\label{eq:spliting variance}
    \t{Var}(J^2)=\sum_{i,j\in\{x,y,z\}}\t{Cov}(J_i^2,J_j^2)
\end{equation}
and express the $\t{Cov}(J_i^2,J_j^2)$ in terms of qubit moments. 

First let us consider $i=j$ and the following moment:
\begin{equation}
\label{eq:definition of indices}
    \langle J_i^4\rangle=\frac{1}{16}\sum_{k,l,m,n=1}^N\langle\sigma_i^{(k)}\sigma_i^{(l)}\sigma_i^{(m)}\sigma_i^{(n)}\rangle.
\end{equation}
Depending on whether some of $k,l,m,n$ indices are equal, the term $\langle\sigma_i^{(k)}\sigma_i^{(l)}\sigma_i^{(m)}\sigma_i^{(n)}\rangle$ may evaluate to different qubit moments, due to $\sigma_i^2=\id$ identity. We list all possible cases and number of their occurrences in the sum \eqref{eq:definition of indices}:
\begin{itemize}
    \item All indices are different, e.g. $(k,l,m,n)=(1,2,3,4)$: $N(N-1)(N-2)(N-3)$
    \item Two indices are equal, e.g. $(k,l,m,n)=(1,1,2,3)$: $6N(N-1)(N-2)$
    \item Three indices are equal, e.g. $(k,l,m,n)=(1,1,1,2)$: $4N(N-1)$
    \item Two pairs of equal indices, e.g. $(k,l,m,n)=(1,1,2,2)$: $3N(N-1)$
    \item Four indices are equal, e.g. $(k,l,m,n)=(1,1,1,1)$: $N$
\end{itemize}
and after collecting the contributions from all cases we get:
\begin{align}
    \langle J_i^4\rangle=\frac{1}{16}[&N^4\langle \sigma_i\otimes\sigma_i\otimes\sigma_i\otimes\sigma_i\rangle+\\+&N^3(6\langle \sigma_i\otimes\sigma_i\rangle-6\langle \sigma_i\otimes\sigma_i\otimes\sigma_i\otimes\sigma_i\rangle)+\nonumber\\+&N^2(3-14\langle \sigma_i\otimes\sigma_i\rangle+11\langle \sigma_i\otimes\sigma_i\otimes\sigma_i\otimes\sigma_i\rangle)+\nonumber\\+&N(-2+8\langle \sigma_i\otimes\sigma_i\rangle - 6\langle \sigma_i\otimes\sigma_i\otimes\sigma_i\otimes\sigma_i\rangle)].\nonumber
\end{align}
We also find
\begin{align}
    \langle J_i^2\rangle^2=\frac{1}{16}[&N^4\langle \sigma_i\otimes\sigma_i\rangle^2+\\+&N^3(2\langle \sigma_i\otimes\sigma_i\rangle-2\langle \sigma_i\otimes\sigma_i\rangle^2)+\nonumber\\+&N^2(1-2\langle \sigma_i\otimes\sigma_i\rangle+\langle \sigma_i\otimes\sigma_i\rangle^2)],\nonumber
\end{align}
and we finally get:
\begin{align}
    &\t{Cov}(J_i^2,J_i^2)=\frac{1}{16}[N^4(\langle \sigma_i\otimes\sigma_i\otimes\sigma_i\otimes\sigma_i\rangle-\langle \sigma_i\otimes\sigma_i\rangle^2)+\nonumber\\&+N^3(4\langle \sigma_i\otimes\sigma_i\rangle+2\langle \sigma_i\otimes\sigma_i\rangle^2-6\langle \sigma_i\otimes\sigma_i\otimes\sigma_i\otimes\sigma_i\rangle)+\nonumber\\&+N^2(2-12\langle \sigma_i\otimes\sigma_i\rangle-\langle \sigma_i\otimes\sigma_i\rangle^2+\nonumber\\&\qquad\qquad\ +11\langle \sigma_i\otimes\sigma_i\otimes\sigma_i\otimes\sigma_i\rangle)+\nonumber\\&+N(-2+8\langle \sigma_i\otimes\sigma_i\rangle - 6\langle \sigma_i\otimes\sigma_i\otimes\sigma_i\otimes\sigma_i\rangle)].
\end{align}
After applying the approximate Gaussian property \eqref{eq:gaussian} it reads:
\begin{align}
    \t{Cov}(J_i^2,J_i^2)=\frac{N^3}{4}[\langle \sigma_i\otimes\sigma_i\rangle&-\langle \sigma_i\otimes\sigma_i\rangle^2 + O(N^3\chi^4)].
\end{align}
We now consider the case $i\neq j$ with the following possible cases of qubit indices in term $\langle\sigma_i^{(k)}\sigma_i^{(l)}\sigma_j^{(m)}\sigma_j^{(n)}\rangle$:
\begin{itemize}
    \item All indices are different, e.g. $(k,l,m,n)=(1,2,3,4)$: $N(N-1)(N-2)(N-3)$
    \item One of indices $k,l$ is equal to one of indices $m,n$, e.g. $(k,l,m,n)=(1,2,1,3)$: $4N(N-1)(N-2)$
    \item Pair or triplet of equal indices with $k=l$, e.g. $(k,l,m,n)=(1,1,2,3)$: $N^2(N-1)$
    \item Pair or triplet of equal indices with $m=n$, e.g. $(k,l,m,n)=(1,2,3,3)$: $N^2(N-1)$
    \item Two pairs of equal indices described by either $k=m\land l=n$ or $k=n\land l=m$, e.g. $(k,l,m,n)=(1,2,1,2)$: $2N(N-1)$
    \item Either two equal pairs or all four indices equal described by $k=l$ and $m=n$, e.g. $(k,l,m,n)=(1,1,2,2)$: $N^2$
\end{itemize}
and collecting all contributions results in:
\begin{align}
    \frac{1}{2}&\langle J_i^2J_j^2+J_j^2J_i^2\rangle=\frac{1}{16}[N^4\langle\sigma_i\otimes\sigma_i\otimes\sigma_j\otimes\sigma_j\rangle+\\&+N^3(\langle \sigma_i\otimes\sigma_i\rangle +\langle \sigma_j\otimes\sigma_j\rangle-6\langle \sigma_i\otimes\sigma_i\otimes\sigma_j\otimes\sigma_j\rangle)+\nonumber\\&+N^2(1-\langle \sigma_i\otimes\sigma_i\rangle-\langle \sigma_j\otimes\sigma_j\rangle-2\langle \sigma_k\otimes\sigma_k\rangle+\nonumber\\&\qquad\qquad\qquad\qquad+11\langle \sigma_i\otimes\sigma_i\otimes\sigma_j\otimes\sigma_j\rangle)+\nonumber\\&+N(2\langle \sigma_k\otimes\sigma_k\rangle-6\langle \sigma_i\otimes\sigma_i\otimes\sigma_j\otimes\sigma_j\rangle)],\nonumber
\end{align}
with $k\in\{x,y,z\}$ being different than $i$ and $j$. We additionally find
\begin{align}
    \langle &J_i^2\rangle\langle J_j^2\rangle=\frac{1}{16}[N^4\langle\sigma_i\otimes\sigma_i\rangle\langle\sigma_j\otimes\sigma_j\rangle +\\&+N^3(\langle\sigma_i\otimes\sigma_i\rangle+\langle\sigma_j\otimes\sigma_j\rangle-2\langle\sigma_i\otimes\sigma_i\rangle\langle\sigma_j\otimes\sigma_j\rangle) +\nonumber\\&+N^2(1-\langle\sigma_i\otimes\sigma_i\rangle-\langle\sigma_j\otimes\sigma_j\rangle+\langle\sigma_i\otimes\sigma_i\rangle\langle\sigma_j\otimes\sigma_j\rangle)]\nonumber
\end{align}
and finally:
\begin{align}
    &\t{Cov}(J_i^2,J_j^2)=\frac{1}{16}[N^4(\langle\sigma_i\otimes\sigma_i\otimes\sigma_j\otimes\sigma_j\rangle+\\&\qquad\qquad\qquad\qquad\quad-\langle\sigma_i\otimes\sigma_i\rangle\langle\sigma_j\otimes\sigma_j\rangle)+\nonumber\\&+N^3(2\langle\sigma_i\otimes\sigma_i\rangle\langle\sigma_j\otimes\sigma_j\rangle-6\langle \sigma_i\otimes\sigma_i\otimes\sigma_j\otimes\sigma_j\rangle)+\nonumber\\&+N^2(-2\langle \sigma_k\otimes\sigma_k\rangle-\langle\sigma_i\otimes\sigma_i\rangle\langle\sigma_j\otimes\sigma_j\rangle+\nonumber\\&\qquad\qquad\qquad\qquad\ +11\langle \sigma_i\otimes\sigma_i\otimes\sigma_j\otimes\sigma_j\rangle)+\nonumber\\&+N(2\langle \sigma_k\otimes\sigma_k\rangle-6\langle \sigma_i\otimes\sigma_i\otimes\sigma_j\otimes\sigma_j\rangle)].\nonumber
\end{align}
The approximate Gaussian property \eqref{eq:gaussian} allows for the following expression:
\begin{align}
    \t{Cov}(J_i^2,J_j^2)=\frac{N^3}{4}[&-\langle\sigma_i\otimes\sigma_i\rangle\langle\sigma_j\otimes\sigma_j\rangle+O(N^3\chi^4)].
\end{align}
We can now find the $\t{Var}(J^2)$ using \eqref{eq:spliting variance}, and it reads:
\begin{align}
    \t{Var}(J^2)&=\frac{N^3}{4}[\eta^2(1-\eta^2)+O(N^3\chi^4)].
\end{align}
\end{document}